      [1999/12/01 v1.4c Il Nuovo Cimento]
\documentclass{cimento}
\begin{document}

\title{Status of the Standard Model at the LHC Start}
\author{G.~Altarelli\from{ins:x}}
\instlist{\inst{ins:x} 
Dipartimento di Fisica `E.~Amaldi', Universit\`a di Roma Tre
and INFN, Sezione di Roma Tre, I-00146 Rome, Italy and \\ CERN, Department of Physics, Theory Unit, 
 CH--1211 Geneva 23, Switzerland}
\maketitle
\begin{abstract}
I present a concise review of where we stand in particle physics today. First, I will discuss QCD, then the electroweak sector and finally the motivations and the avenues for new physics beyond the Standard Model.

\end{abstract}
\begin{flushright}
{RM3-TH/08-8}
{CERN-PH-TH/2008-084}\\
\end{flushright}

\section{QCD}
\label{sec:1}

QCD stands as a main building block of the Standard Model (SM) of particle physics. For many years the relativistic quantum field theory of reference was QED, but at present QCD offers a more complex and intriguing theoretical laboratory. Indeed, due to asymptotic freedom, QCD can be considered as a better defined theory than QED. The statement that QCD is an unbroken renormalizable gauge theory with six kinds of triplets quarks with given masses completely specifies the form of the Lagrangian in terms of quark and gluon fields. From the compact form of its Lagrangian one might be led to  think that QCD is a "simple" theory. But actually this simple theory has an extremely rich dynamical content, including the striking properties of asymptotic freedom and of confinement,  the complexity of the observed hadronic spectrum (with light and heavy quarks), the spontaneous breaking of (approximate) chiral symmetry, a complicated phase transition structure (deconfinement, chiral symmetry restoration, colour superconductivity), a highly non trivial vacuum topology (instantons, $U(1)_A$ symmetry breaking, strong CP violation,....), and so on.

How do we get predictions from QCD? There are non perturbative methods: lattice simulations (in great continuous progress), effective lagrangians valid in restricted specified domains [chiral lagrangians, heavy quark effective theories, Soft Collinear Effective Theories (SCET), Non Relativistic QCD....] and also QCD sum rules, potential models (for quarkonium) etc. But the perturbative approach, based on asymptotic freedom and valid for hard processes, still remains the main quantitative connection to experiment.

Due to confinement no free coloured particles are observed but only colour singlet hadrons. In high energy collisions the produced quarks and gluons materialize as narrow jets of hadrons. Our understanding of the confinement mechanism has much improved thanks to lattice simulations of QCD at finite temperatures and densities \cite{ref:ej}.  The potential between two colour charges clearly shows a linear slope at large distances (linearly rising potential). The slope decreases with increasing temperature until it vanishes at a critical temperature $T_C$. Above $T_C$ the slope remains zero. The phase transitions of colour deconfinement and of chiral restoration appear to happen together on the lattice. Near the critical temperature for both deconfinement and chiral restoration a rapid transition is observed in lattice simulations. In particular the energy density $\epsilon(T)$ is seen to sharply increase. The critical parameters and the nature of the phase transition depend on the number of quark flavours $N_f$ and on their masses. For example, for  $N_f$ = 2 or 2+1 (i.e. 2 light u and d quarks and 1 heavier s quark), $T_C \sim 175~MeV$  and $\epsilon(T_C) \sim 0.5-1.0 ~GeV/fm^3$. For realistic values of the masses $m_s$ and $m_{u,d}$ the phase transition appears to be a second order one, while it becomes first order for very small or very large $m_{u,d,s}$. The hadronic phase and the deconfined phase are separated by a crossover line at small densities and by a critical line at high densities. Determining the exact location of the critical point in T and $\mu_B$ is an important challenge for theory and is also important for the interpretation of heavy ion collision experiments. At high densities the colour superconducting phase is probably also present with bosonic diquarks acting as Cooper pairs. 

A large investment is being done in experiments of heavy ion collisions with the aim of finding some evidence of the quark gluon plasma phase. Many exciting results have been found at the CERN SPS in the past  years and more recently at RHIC \cite{ref:rhi}. At the CERN SPS some experimental hints of variation with the energy density were found in the form, for example, of $J/ \Psi$ production suppression or of strangeness enhancement when going from p-A to Pb-Pb collisions. Indeed a posteriori the CERN SPS appears well positioned in energy to probe the transition region, in that a marked variation of different observables was observed. The most impressive effect detected at RHIC, interpreted as due to the formation of a hot and dense bubble of matter, is the observation of a strong suppression of back-to-back correlations in jets from central collisions in Au-Au, showing that the jet that crosses the bulk of the dense region is absorbed. The produced hot matter shows a high degree of collectivity, as shown by the observation of elliptic flow (produced hadrons show an elliptic distribution while it would be spherical for a gas) and resembles a perfect liquid with small or no viscosity. There is also evidence for a 2-component hadronisation mechanism: 
coalescence \cite{ref:coa} and fragmentation. Early produced partons with high density show an exponential falling in $p_T$: they produce hadrons by joining together. At large $p_T$ fragmentation with power behaviour survives. Elliptic flow, inclusive spectra, partonic energy loss in medium, strangeness enhancement, J/$\Psi$  suppression etc. are all suggestive (but only suggestive!) of early production of a coloured partonic medium with high energy density
and temperature, close to the theoretically expected values, then expanding as a near ideal fluid. The experimental programme on heavy ion collisions will continue at RHIC and at the LHC where ALICE, the dedicated heavy ion collision experiment is progressing towards commissioning.

As we have seen, a main approach to non perturbative problems in QCD is by simulations of the theory on the lattice, a technique started by K. Wilson in 1974 which has shown continuous progress over the last decades. One recent big step, made possible by the availability of more powerful dedicated computers, is the evolution from quenched (i.e. with no dynamical fermions) to unquenched calculations. In doing so an evident improvement in the agreement of predictions with the data is obtained. For example \cite{ref:kro}, modern unquenched simulations reproduce the hadron spectrum quite well. Calculations with dynamical fermions (which take into account the effects of virtual quark loops) imply the evaluation of the quark determinant which is a difficult task. How difficult depends on the particular calculation method. There are several approaches (Wilson, twisted mass,  Kogut-Susskind staggered, Ginsparg-Wilson fermions), each with its own advantages and disadvantages (including the time it takes to run the simulation on a computer). Another area of progress is the implementation of chiral extrapolations: lattice simulation is limited to large enough masses of light quarks. To extrapolate the results down to the physical pion mass one can take advantage of the chiral effective theory in order to control the chiral logs: $\log(m_q/4\pi f_\pi)$. For lattice QCD one is now in an epoch of pre-dictivity as opposed to the  post-dictivity of the past. And in fact the range of precise lattice results currently includes many domains:  the QCD coupling constant (the value $\alpha_s(m_Z)=0.1170(12)$ has been recently quoted \cite{ref:alf}: the central value is in agreement with other determinations but I would not trust the stated error as a fair representation of the total uncertainty), the quark masses, the form factors for K and D decay, the B parameter for kaons, the decay constants $f_K$, $f_D$, $f_{Ds}$, the $B_c$ mass and many more.

We  now discuss  perturbative QCD. In the QCD Lagrangian quark masses are the only parameters with dimensions. Naively (or classically) one would expect massless QCD to be scale invariant so that dimensionless observables would not depend on the absolute energy scale but only on ratios of energy variables. While massless QCD in the quantum version, after regularisation and renormalisation, is finally not scale invariant, the theory is asymptotically free and all the departures from scaling are asymptotically small, logarithmic and computable in terms of the running coupling $\alpha_s(Q^2)$. Mass corrections, present in the realistic case together with non perturbative effects, are suppressed by powers. The QCD beta function that fixes the running coupling is known in QCD up to 4 loops in the $MS$ or $\bar{MS}$ definitions and the expansion is well behaved. The 4-loop calculation  \cite{ref:van} involving about 50.000 4-loop diagrams is a great piece of work. The running coupling is a function of $\log{Q^2/\Lambda^2_{QCD}}$, where $\Lambda_{QCD}$ is the scale that breaks scale invariance in massless QCD. Its value in $\bar{MS}$, for 5 flavours of quarks, from the PDG'06 is $\Lambda_{QCD}\sim 222(25)~MeV$. There is no hierarchy problem in QCD, in that the logarithmic evolution of the running makes the smallness of $\Lambda_{QCD}$ with respect to the Planck mass $M_{Pl}$ natural: $\Lambda_{QCD}\sim M_{Pl} \exp{[-1/2b\alpha_s(M_{Pl}^2)]}$. 

The measurements of $\alpha_s(Q^2)$  are among the main quantitative tests of the theory. The most  precise and  reliable determinations are from $e^+e^-$ colliders (mainly at LEP: inclusive Z decays, inclusive
hadronic $\tau$ decay, event shapes and jet rates) and from scaling violations in Deep Inelastic Scattering (DIS).  There is a remarkable agreement among the different determinations. If I directly average the best values from inclusive Z decays, from $R_\tau$, from event shapes and jet rates in $e^+e^-$, from $F_3$ and from $F_2$ in DIS I obtain \cite{ref:alt} $\alpha_s(m_Z^2)=0.1187(16)$ in good agreement with the PDG'06 average $\alpha_s(m_Z^2)=0.1176(20)$.

Since $\alpha_s$ is not too small, $\alpha_s(m_Z^2) \sim 0.12$, the need of high
order perturbative calculations, resummation of logs at all 
orders etc. is particularly acute.  Ingenious new computational techniques and software have been developed and many calculations have been realized that only a decade ago appeared as impossible. Some examples follow.

In 2004 the complete calculation of the NNLO splitting functions has been published \cite{ref:moc} $\alpha_s P \sim \alpha_s P_1+ \alpha_s^2 P_2 + \alpha_s^3 P_3+\dots$, a really monumental, fully analytic, computation. More recently the main part of the inclusive hadronic $Z$ and $\tau$ decays at $o(\alpha_s^4)$ (NNNLO!) has been computed \cite{ref:bck}. The calculation (which involves  some 20.000 diagrams) is complete for $\tau$ decay, while for $Z$ decay only the non singlet terms, proportional to $\Sigma_f Q_f^2$, are included
( but singlet terms ~$(\Sigma_f Q_f)^2)$ are small at the previous order $o(\alpha_s^3)$). The result, for $n_f=5$ and $a_s= \alpha_s(m_Z^2)/\pi$  (relevant for $Z$ decay), is $R(Q^2)=3 \Sigma_f Q_f^2 [1+a_s +1.4097a_s^2 -12.76709a_s^3 - 80.0075a_s^4+...]$. This result is important to improve the derivation of $\alpha_s$ from hadronic $Z$ and $\tau$ decays. The authors point out that the new corrections bring the two central values closer together, with the values: $\alpha_s(m_Z^2)|_Z= 0.1190(26)$ and $\alpha_s(m_Z^2)|_\tau= 0.1202(19)$. The calculation of the hadronic event shapes in $e^+e^-$ annihilation at NNLO has also been completed \cite{ref:ggh}, which involves consideration of 3, 4 and 5 jets with one loop corrections for 4 jets and two loop corrections for 3 jets. These calculations were applied in ref.\cite{ref:dis} to the measurement of $\alpha_s$ from data on event shapes obtained by ALEPH with the result: $\alpha_s(m_Z^2)=0.1240(34)$. 

The importance of DIS for QCD goes well beyond the measurement of $\alpha_s$. In the past it played a crucial role in establishing the reality of quarks and gluons as partons and in promoting  QCD as the theory of strong interactions. Nowadays it still generates challenges to QCD as, for example, in the domain of structure functions at small x or of polarized structure functions or of generalized parton densities and so on. 

The problem of constructing a convergent procedure to include the BFKL corrections at small x in the singlet splitting functions, in agreement with the small-x behaviour observed at HERA, has been a long standing puzzle which has now been essentially solved. The naive BFKL rise of splitting functions is tamed by resummation of collinear singularities and by running coupling effects. The resummed expansion is well behaved and the result is close to the perturbative NLO splitting function in the region of HERA data at small x \cite{ref:abf},\cite{ref:ccss}. 

In polarized DIS one main question is how the proton helicity is distributed among quarks, gluons and orbital angular momentum: $1/2\Delta \Sigma + \Delta g + L_z= 1/2$.
The quark moment $\Delta \Sigma$ was found to be small: typically, at $Q^2\sim 1 GeV^2$, $\Delta \Sigma_{exp} \sim 0.3$ (the "spin crisis")\cite{ref:spi1}. Either $\Delta g + L_z$ is large or there are contributions to $\Delta \Sigma$ at very small x outside of the measured region. $\Delta g$ evolves like $\Delta g \sim log Q^2$, so that eventually should become large (while $\Delta \Sigma$ and $\Delta g + L_z$ are $Q^2$ independent in LO). It will take long before this log growth of $\Delta g$ will be confirmed by experiment! $\Delta g$ can be measured indirectly by scaling violations and directly from asymmetries, e.g. in $c \bar c$ production. Existing direct measurements by Hermes, Compass, and at RHIC are still very crude and show no hint of a large $\Delta g$ \cite{ref:spi2}.

Another important role of DIS is to provide information on parton density functions (PDF) which are instrumental for computing cross-sections of hard processes at hadron colliders via the factorisation formula. The predictions for cross sections and distributions at $pp$ or $p\bar p$ colliders for large $p_T$ jets or photons, for heavy quark production, for Drell-Yan, W and Z production are all in very good agreement with experiment. There was an apparent problem for b quark production at the Tevatron, but the problem appears now to be solved by a combination of refinements (log resummation, B hadrons instead of b quarks, better fragmentation functions....)\cite{ref:cac}. The QCD predictions are so solid that W and Z production are actually considered as possible luminosity monitors for the LHC. 

A great effort is being devoted to the preparation to the LHC. Calculations for specific processes are being completed. A very important example is Higgs production via $g ~+~ g \rightarrow H$ \cite{ref:boz}. The amplitude is dominated by the top quark loop (if heavier coloured particles exist they also would contribute). The NLO corrections turn out to be particularly large. Higher order corrections can be computed either in the effective lagrangian approach, where the heavy top is integrated away and the loop is shrunk down to a point (the coefficient of the effective vertex is known to $\alpha_s^4$ accuracy), or in the full theory. At the NLO the two approaches agree very well for the rate as a function of $m_H$. The NNLO corrections have been computed in the effective vertex approximation. Beyond fixed order, resummation of large logs were carried out. Also the NLO EW contributions are known by now. Rapidity (at  NNLO) and $p_T$ distributions (at NLO) have also been evaluated. At smaller $p_T$ the large logarithms $[log(p_T/m_H)]^n$ have been resummed in analogy with what was done long ago for W and Z production. 

The activity on event simulation also received a big boost from the LHC preparation (see, for example, the review \cite{ref:LHC}). General algorithms for performing NLO calculations numerically (requiring techniques for the cancellation of singularities between real and virtual diagrams) have been developed (see, for example, \cite{ref:num}). The matching of matrix element calculation of rates together with the modeling of parton showers has been realised in packages, as for example in the MC@NLO \cite{ref:frix} or POWHEG \cite{ref:fnr} based on HERWIG. The matrix element calculation, improved by resummation of large logs, provides the hard skeleton (with large $p_T$ branchings) while the parton shower is constructed by a sequence of factorized collinear emissions fixed by the QCD splitting functions. In addition, at low scales a model of hadronisation completes the simulation. The importance of all the components, matrix element, parton shower and hadronisation can be appreciated in simulations of hard events compared with the Tevatron data. 

In conclusion, I think that the domain of QCD appears as one of great maturity but also of robust vitality and all the QCD predictions that one was able to formulate and to test are in very good agreement with experiment.

\section{The Higgs Problem}
\label{sec:4}

The Higgs problem is really central in particle physics today. On the one hand, the experimental verification of the Standard Model (SM) cannot be considered complete until the structure of the  Higgs sector is not established by experiment. On the other hand, the Higgs is directly related to most of the major open problems of particle physics, like the flavour problem or the hierarchy problem, the latter strongly suggesting the need for new physics near the weak scale, which could also clarify  the dark matter identity. It is clear that the fact that some sort of Higgs mechanism is at work has already been established. The W or the Z with longitudinal polarization that we observe are not present in an unbroken gauge theory (massless spin-1 particles, like the photon, are transversely polarized). The longitudinal degree of freedom for the W or the Z is borrowed from the Higgs sector and is an evidence for it. Also, the couplings of quarks and leptons to
the weak gauge bosons W$^{\pm}$ and Z are indeed precisely those
prescribed by the gauge symmetry.  To a lesser
accuracy the triple gauge vertices $\gamma$WW and ZWW have also
been found in agreement with the specific predictions of the
$SU(2)\bigotimes U(1)$ gauge theory. This means that it has been
verified that the gauge symmetry is unbroken in the vertices of the
theory: all currents and charges are indeed symmetric. Yet there is obvious
evidence that the symmetry is instead badly broken in the
masses. Not only the W and the Z have large masses, but the large splitting of, for example,  the t-b doublet shows that even a global weak SU(2) is not at all respected by the fermion spectrum. This is a clear signal of spontaneous
symmetry breaking and its implementation in a gauge theory is via the Higgs mechanism. The big remaining questions are about
the nature and the properties of the Higgs particle(s). 

The LHC has been designed to solve the Higgs problem. It is well known that in the SM with only one Higgs doublet a lower limit on
$m_H$ can be derived from the requirement of vacuum stability (i.e. that the quartic Higgs coupling $\lambda$ does not turn negative in its running up to a large scale $\Lambda$) or, in milder form, of a moderate instability, compatible with the lifetime of the Universe  \cite{ref:isid}. The Higgs mass enters because it fixes the initial value of the quartic Higgs coupling $\lambda$. For the actual value of $m_t$ the lower limit is below the direct experimental bound for $\Lambda \sim $ a few TeV and is $M_H> 130$ GeV for $\Lambda \sim M_{Pl}$. Similarly an upper bound on $m_H$ (with mild dependence
on $m_t$) is obtained, as described in \cite{ref:hri}, from the requirement that for $\lambda$ no Landau pole appears up to the scale $\Lambda$, or in simpler terms, that the perturbative description of the theory remains valid up to  $\Lambda$. The upper limit on the Higgs mass in the SM is clearly important for assessing the chances of success of the LHC as an accelerator designed to solve the Higgs problem. Even if 
$\Lambda$ is as small as ~a few TeV the limit is $m_H < 600-800~$GeV and becomes $m_H < 180~$GeV for
$\Lambda \sim M_{Pl}$. An additional argument indicating that the solution of the Higgs problem cannot be too far away is the fact that, in the absence of a Higgs particle or of an alternative mechanism, violations of unitarity appear in scattering amplitudes involving longitudinal gauge bosons (those most directly related to the Higgs sector) at energies in the few TeV range \cite{ref:unit}. In conclusion, it is very unlikely that the solution of the Higgs problem can be missed at the LHC.

\section{Precision Tests of the Standard Electroweak Theory}
\label{sec:5}

The most precise tests of the electroweak theory apply to the QED sector. The anomalous magnetic moments of the electron and of the muon are among the most precise measurements in the whole of physics. Recently there have been new precise measurements of $a_e$ and $a_\mu$ for the electron \cite{ref:ae1} and the muon \cite{ref:amu} ($a = (g-2)/2$).The QED part has been computed analytically for $i=1,2,3$, while for $i=4$ there is a numerical calculation (see, for example, \cite{ref:kino}). Some terms for $i=5$ have also been estimated for the muon case. The weak contribution is from $W$ or $Z$ exchange. The hadronic contribution is from vacuum polarization insertions and from light by light scattering diagrams.  For the electron case the weak contribution is essentially negligible and the hadronic term does not introduce an important uncertainty.  As a result the $a_e$ measurement can be used to obtain the most precise determination of the fine structure constant \cite{ref:ae2}. In the muon case the experimental precision is less by about 3 orders of magnitude, but the sensitivity to new physics effects is typically increased by a factor $(m_\mu/m_e)^2 \sim 4^.10^4$. The dominant theoretical ambiguities arise from the hadronic terms in vacuum polarization and in light by light scattering. If the vacuum polarization terms are evaluated from the $e^+e^-$ data a discrepancy of $\sim 3 \sigma$ is obtained (the $\tau$ data would indicate better agreement, but the connection to $a_\mu$ is less direct and recent new data have added solidity to the $e^+e^-$ route)\cite{ref:amu2}. Finally, we note that, given the great accuracy of the $a_\mu$ measurement and the estimated size of the new physics contributions, for example from SUSY, it is not unreasonable that a first signal of new physics would appear in this quantity.

The results of the electroweak precision tests also including the measurements of $m_t$, $m_W$ and the searches for new physics at the Tevatron form a very stringent set of precise constraints \cite{ref:ewg} to compare with the Standard Model (SM) or with
any of its conceivable extensions. When confronted with these results, on the whole the SM performs rather
well, so that it is fair to say that no clear indication for new physics emerges from the data \cite{ref:AG}.  But the
Higgs sector of the SM is still very much untested. What has been
tested is the relation $M_W^2=M_Z^2\cos^2{\theta_W}$, modified by small, computable
radiative corrections. This relation means that the effective Higgs
(be it fundamental or composite) is indeed a weak isospin doublet.
The Higgs particle has not been found but in the SM its mass can well
be larger than the present direct lower limit $m_H > 114.4$~GeV
obtained from direct searches at LEP-2.  The radiative corrections
computed in the SM when compared to the data on precision electroweak
tests lead to a clear indication for a light Higgs, not too far from
the present lower bound. The exact upper limit for $m_H$ in the SM depends on the value of the top quark mass $m_t$ (the one-loop radiative corrections are quadratic in $m_t$ and logarithmic in $m_H$). The measured value of $m_t$ went down recently (as well as the associated error) according to the results of Run II at the Tevatron. The CDF and D0 combined value is at present $m_t~= 172.6~\pm~1.4~GeV$. As a consequence the present limit on $m_H$ is quite stringent: $m_H < 190~GeV$ (at $95\%$ c.l., after including the information from the 114.4 GeV direct bound)  \cite{ref:ewg}.  

\section{The Physics of Flavour}
\label{sec:3}

In the last decade great progress in different areas of flavour physics has been achieved. In the quark sector, the amazing results of a generation of frontier experiments, obtained at B factories and at accelerators, have become available. QCD has been playing a crucial role in the interpretation of experiments by a combination of effective theory methods (heavy quark effective theory, NRQCD, SCET), lattice simulations and perturbative calculations. A great achievement obtained by many theorists over the last years is the calculation at NNLO of the branching ratio for $B\rightarrow X_s \gamma$ with B a beauty meson \cite{ref:mis}. The effect of the photon energy cut, $E_\gamma > E_0$, necessary in practice, has been evaluated at NNLO \cite{ref:bec}. The central value of the theoretical prediction is now slightly below the data: for $B[B\rightarrow X_s\gamma, E_0=1.6~GeV](10^{-4})$ the experimental value is 3.55(26) and the theoretical value is 3.15(23) \cite{ref:mis} or 2.98(26) \cite{ref:bec}, which to me is good agreement.
The hope of the B-decay experiments was to detect departures from the CKM picture of mixing and of CP violation as  signals of new physics. Finally, B mixing and CP violation agree very well with the SM predictions based on the CKM matrix \cite{ref:fle}. The recent measurement of $\Delta m_s$ by CDF and D0, in fair agreement with the SM expectation, has closed another door for new physics. But in some channels, especially those which occur through penguin loops, it is well possible that substantial deviations could be hidden (possible hints are reported in $B\rightarrow K\pi$ decays \cite{ref:nat} and in $b\rightarrow s$ transitions \cite{ref:ciu}). But certainly the amazing performance of the SM in flavour changing  and/or CP violating transitions in K and B decays poses very strong constraints on all proposed models of new physics \cite{ref:isi}. 

In the leptonic sector the study of neutrino oscillations has led to the discovery that at least two neutrinos are not massless and to the determination of the mixing matrix \cite{ref:alfe}. Neutrinos are not all massless but their masses are very small (at most a fraction of $eV$). Probably masses are small because $\nu$Õs are Majorana fermions, and, by the see-saw mechanism, their masses are inversely proportional to the large scale $M$ where lepton number ($L$) non conservation occurs (as expected in GUT's). Indeed the value of $M\sim m_{\nu R}$ from experiment is compatible with being close to $M_{GUT} \sim 10^{14}-10^{15}GeV$, so that neutrino masses fit well in the GUT picture and actually support it. The interpretation of neutrinos as Majorana particles enhances the importance of experiments aimed at the detection of neutrinoless double beta decay and a huge effort in this direction is underway.  It was realized that decays of heavy $\nu_R$ with CP and L non conservation can produce a B-L asymmetry. The range of neutrino masses indicated by neutrino phenomenology turns out to be perfectly compatible with the idea of baryogenesis via leptogenesis \cite{ref:buch}. This elegant model for baryogenesis has by now replaced the idea of baryogenesis near the weak scale, which has been strongly disfavoured by LEP. It is remarkable that we now know the neutrino mixing matrix with good accuracy. Two mixing angles are large and one is small. The atmospheric angle $\theta_{23}$ is large, actually compatible with maximal but not necessarily so: at $3\sigma$ \cite{ref:mal}: $0.34 \leq \sin^2{\theta_{23}}\leq 0.68$ with central value around  $0.5$. The solar angle $\theta_{12}$ (the best measured) is large, $\sin^2{\theta_{12}}\sim 0.3$, but certainly not maximal (by more than 5$\sigma$). The third angle $\theta_{13}$, strongly limited mainly by the CHOOZ experiment, has at present a $3\sigma$ upper limit given by about $\sin^2{\theta_{13}}\leq 0.04$. The non conservation of the three separate lepton numbers and the large leptonic mixing angles make it possible that processes like $\mu \rightarrow e \gamma$ or $\tau \rightarrow \mu \gamma$ might be observable, not in the SM but in extensions of it like the MSSM. Thus, for example, the outcome of the now running experiment MEG at PSI \cite{ref:meg}, aiming at improving the limit on $\mu \rightarrow e \gamma$ by 1 or 2 orders of magnitude, is of great interest. 

\section{Outlook on Avenues beyond the Standard Model}
\label{sec:5}

No signal of new physics has been
found neither in electroweak precision tests nor in flavour physics. Given the success of the SM why are we not satisfied with that theory? Why not just find the Higgs particle,
for completeness, and declare that particle physics is closed? The reason is that there are
both conceptual problems and phenomenological indications for physics beyond the SM. On the conceptual side the most
obvious problems are the proliferation of parameters, the puzzles of family replication and of flavour hierarchies, the fact that quantum gravity is not included in the SM and the related hierarchy problem. Among the main
phenomenological hints for new physics we can list the quest for Grand Unification and coupling constant merging, dark matter, neutrino masses (explained in terms of L non conservation), 
baryogenesis and the cosmological vacuum energy (a gigantic naturalness problem).

The computed evolution with energy
of the effective gauge couplings clearly points towards the unification of the electro-weak and strong forces (Grand
Unified Theories: GUT's) at scales of energy
$M_{GUT}\sim  10^{15}-10^{16}~ GeV$ which are close to the scale of quantum gravity, $M_{Pl}\sim 10^{19}~ GeV$.  One is led to
imagine  a unified theory of all interactions also including gravity (at present superstrings provide the best attempt at such
a theory). Thus GUT's and the realm of quantum gravity set a very distant energy horizon that modern particle theory cannot
ignore. Can the SM without new physics be valid up to such large energies? One can imagine that some obvious problems could be postponed to the more fundamental theory at the Planck mass. For example, the explanation of the three generations of fermions and the understanding of fermion masses and mixing angles can be postponed. But other problems must find their solution in the low energy theory. In particular, the structure of the
SM could not naturally explain the relative smallness of the weak scale of mass, set by the Higgs mechanism at $\mu\sim
1/\sqrt{G_F}\sim  250~ GeV$  with $G_F$ being the Fermi coupling constant. This so-called hierarchy problem is due to the instability of the SM with respect to quantum corrections. This is related to
the
presence of fundamental scalar fields in the theory with quadratic mass divergences and no protective extra symmetry at
$\mu=0$. For fermion masses, first, the divergences are logarithmic and, second, they are forbidden by the $SU(2)\bigotimes
U(1)$ gauge symmetry plus the fact that at
$m=0$ an additional symmetry, i.e. chiral  symmetry, is restored. Here, when talking of divergences, we are not
worried of actual infinities. The theory is renormalizable and finite once the dependence on the cut off $\Lambda$ is
absorbed in a redefinition of masses and couplings. Rather the hierarchy problem is one of naturalness. We can look at the
cut off as a parameterization of our ignorance on the new physics that will modify the theory at large energy
scales. Then it is relevant to look at the dependence of physical quantities on the cut off and to demand that no
unexplained enormously accurate cancellations arise. 

The hierarchy problem can be put in less abstract terms: loop corrections to the higgs mass squared are
quadratic in the cut off $\Lambda$. The most pressing problem is from the top loop.
 With $m_h^2=m^2_{bare}+\delta m_h^2$ the top loop gives 
 \begin{eqnarray}
\delta m_{h|top}^2\sim -\frac{3G_F}{2\sqrt{2} \pi^2} m_t^2 \Lambda^2\sim -(0.2\Lambda)^2 \label{top}
\end{eqnarray}
If we demand that the correction does not exceed the light Higgs mass indicated by the precision tests, $\Lambda$ must be
close, $\Lambda\sim o(1~TeV)$. Similar constraints arise from the quadratic $\Lambda$ dependence of loops with gauge bosons and
scalars, which, however, lead to less pressing bounds. So the hierarchy problem demands new physics to be very close (in
particular the mechanism that quenches the top loop). Actually, this new physics must be rather special, because it must be
very close, yet its effects are not clearly visible (the "LEP Paradox" \cite{ref:BS}) now also accompanied by a similar "flavour paradox" \cite{ref:isi}. Examples of proposed classes of solutions
for the hierarchy problem are:

¥ $\bf{Supersymmetry.}$ In the limit of exact boson-fermion symmetry \cite{ref:Martin} the quadratic divergences of bosons cancel so that
only log divergences remain. However, exact SUSY is clearly unrealistic. For approximate SUSY (with soft breaking terms),
which is the basis for all practical models, $\Lambda$ is replaced by the splitting of SUSY multiplets, $\Lambda\sim
m_{SUSY}-m_{ord}$. In particular, the top loop is quenched by partial cancellation with s-top exchange, so the s-top cannot be too heavy. An important phenomenological indication is that coupling unification is quantitatively precise in SUSY GUT's and that proton decay bounds are not in contradiction with the predictions. An interesting exercise is to repeat the fit of precision tests in the Minimal Supersymmetric Standard Model with GUT constraints added, also including the additional data on the muon $(g-2)$, the dark matter relic density and on the $b\rightarrow s \gamma$ rate. The result is that the central value of the lightest Higgs mass $m_h$ goes up (in better harmony with the bound from direct searches) for moderately large $tan\beta$ and relatively light SUSY spectrum \cite{ref:sus}.

¥ $\bf{Technicolor.}$ The Higgs system is a condensate of new fermions. There are no fundamental scalar Higgs sector, hence no
quadratic devergences associated to the $\mu^2$ mass in the scalar potential. This mechanism needs a very strong binding force,
$\Lambda_{TC}\sim 10^3~\Lambda_{QCD}$. It is  difficult to arrange that such nearby strong force is not showing up in
precision tests. Hence this class of models has been disfavoured by LEP, although some special class of models have been devised a posteriori, like walking TC, top-color assisted TC etc \cite{ref:L-C}. 

¥ $\bf{Extra~dimensions.}$ The idea is that $M_{Pl}$ appears very large, or equivalently that gravity appears very weak,
because we are fooled by hidden extra dimensions so that the real gravity scale is reduced down to a lower scale, even possibly down to
$o(1~TeV)$ ("large" extra dimensions). This possibility is very exciting in itself and it is really remarkable that it is not directly incompatible with experiment but a realistic model has not emerged \cite{ref:Jo}. The most promising set of extra dimensional models are those with "warped" metric, which offer attractive solutions to the hierarchy problem \cite{ref:RS,ref:sura}. An important direction of development is the study of symmetry breaking by orbifolding and/or boundary conditions. These are models where a larger gauge symmetry (with or without SUSY) holds in the bulk. The symmetry is reduced on the 4 dimensional brane, where the physics that we observe is located, as an effect of symmetry breaking induced geometrically by suitable boundary conditions (see, for example, the class of models in \cite{ref:co}). Also "Higgsless  models" have been tried where it is the SM electroweak gauge symmetry which is broken at the boundaries \cite{ref:Hless} (then no Higgs should be found at the LHC but other signals, like additional vector bosons, should appear). Extra dimensions offer a rich and exciting general framework.

¥ $\bf{"Little~Higgs"~models.}$ In these models extra symmetries allow $m_h\not= 0$ only at two-loop level, so that $\Lambda$
can be as large as
$o(10~TeV)$ with the Higgs within present bounds (the top loop is quenched by exchange of heavy vectorlike new  quarks with charge 2/3) \cite{ref:schm}. Certainly these models involve a remarkable level of group theoretic virtuosity. However, in
the simplest versions one is faced with problems with precision tests of the SM . These bad features can be fixed by some suitable complication of the model (see for example, \cite{ref:Ch}). But, in my opinion, the real limit of
this approach is that it only offers a postponement of the main problem by a few TeV, paid by a complete loss of
predictivity at higher energies. In particular all connections to GUT's are lost.

¥ $\bf{Effective~theories~for~compositeness.}$  In this approach \cite{ref:comp} a low energy theory from truncation of some UV completion is described in terms of an elementary sector (the SM particles minus the Higgs) a composite sector (including the Higgs, massive vector bosons $\rho_\mu$ and new fermions) and a mixing sector. The Higgs is a pseudo Goldstone bosons of a larger broken gauge group, with $\rho_\mu$ the corresponding massive vector bosons. Mass eigenstates are mixtures of elementary and composite states, with light particles mostly elementary and heavy particles mostly composite. But the Higgs is totally composite (perhaps also the right-handed top quark). New physics in the composite sector is well hidden because light particles have small mixing angles. The Higgs is light because only acquires
mass through interactions with the light particles from their composite components. This general description can apply to models with a strongly interacting sector as arising from little Higgs or extra dimension scenarios.

¥ $\bf{The~anthropic~solution.}$ The apparent value of the cosmological constant $\Lambda$ poses a tremendous, unsolved naturalness problem \cite{ref:tu}. Yet the value of $\Lambda$ is close to the Weinberg upper bound for galaxy formation \cite{ref:We}. Possibly our Universe is just one of infinitely many (Multiverse) continuously created from the vacuum by quantum fluctuations. Different physics in different Universes according to the multitude of string theory solutions (~$10^{500}$). Perhaps we live in a very unlikely Universe but the only one that allows our existence \cite{ref:anto}. I find applying the anthropic principle to the SM hierarchy problem excessive. After all we can find plenty of models that easily reduce the fine tuning from $10^{14}$ to $10^2$: why make our Universe so terribly unlikely? By comparison the case of the cosmological constant is a lot different: the context is not as fully specified as the for the SM (quantum gravity, string cosmology, branes in extra dimensions, wormholes through different Universes....)

\section{Summary and Conclusion}
\label{sec:6}

Supersymmetry remains the standard way beyond the SM. What is unique to SUSY, beyond leading to a set of consistent and
completely formulated models, as, for example, the MSSM, is that this theory can potentially work up to the GUT energy scale.
In this respect it is the most ambitious model because it describes a computable framework that could be valid all the way
up to the vicinity of the Planck mass. The SUSY models are perfectly compatible with GUT's and are actually quantitatively
supported by coupling unification and compatible with proton decay bounds and also by what we have recently learned on neutrino masses. All other main ideas for going
beyond the SM do not share this synthesis with GUT's. The SUSY way is testable, for example at the LHC, and the issue
of its validity will be decided by experiment. It is true that we could have expected the first signals of SUSY already at
LEP2, based on naturality arguments applied to the most minimal models (for example, those with gaugino universality at
asymptotic scales). The absence of signals has stimulated the development of new ideas like those of extra dimensions
and "little Higgs" models. These ideas are very interesting and provide an important reference for the preparation of LHC
experiments. Models along these new ideas are not so completely formulated and studied as for SUSY and no well defined and
realistic baseline has sofar emerged. But it is well possible that they might represent at least a part of the truth and it
is very important to continue the exploration of new ways beyond the SM. New input from experiment is badly needed, so we all look forward to the start of the LHC.

\acknowledgments
I conclude by thanking the Organisers of this very inspiring Meeting, in particular Lorenzo Foa' and Giancarlo Mantovani for their kind invitation and great hospitality in Perugia. We recognize that this work has been partly supported by the Italian Ministero dell'Universita' e della Ricerca Scientifica, under the COFIN program for 2007-08.
\\
\\
Disclaimer: the list of references is by far incomplete and only meant to provide the reader with a few keys to the literature.

\end{document}